\documentclass[twocolumn, amsmath, amssymb, aps, prl, floatfix, superscriptaddress
]{revtex4-2}
\usepackage{soul}
\usepackage[T1]{fontenc}
\usepackage{graphicx}
\usepackage{bm}
\usepackage{hyperref}
\usepackage{physics}
\usepackage{color}
\usepackage{multirow}
\usepackage{orcidlink}

\renewcommand{\vb}[1]{\boldsymbol{\mathbf{#1}}}
\newcommand{\Cdag}[1]{c^\dagger_{#1}}
\newcommand{\C}[1]{c^{\phantom{\dagger}}_{#1}}

\allowdisplaybreaks

\begin{document}

\title{Topological phonons in anomalous Hall crystals}

\author{Mark R. Hirsbrunner \orcidlink{0000-0001-8115-6098}}
\email{mark.hirsbrunner@utoronto.ca}
\affiliation{
Department of Physics, University of Toronto, Toronto, Ontario M5S 1A7, Canada
}

\author{F\'elix Desrochers \orcidlink{0000-0003-1211-901X}}
\affiliation{
Department of Physics, University of Toronto, Toronto, Ontario M5S 1A7, Canada
}
\affiliation{
Department of Physics, Harvard University, Cambridge, MA 02138, United States}

\author{Joe Huxford \orcidlink{0000-0002-4857-0091}}
\affiliation{
Department of Physics, University of Toronto, Toronto, Ontario M5S 1A7, Canada
}
\affiliation{
Department of Physics and Astronomy, University of Manchester, Oxford Road, Manchester M13 9PL, United Kingdom
}

\author{Yong Baek Kim}
\email{yongbaek.kim@utoronto.ca}
\affiliation{
 Department of Physics, University of Toronto, Toronto, Ontario M5S 1A7, Canada
}

\date{\today}

\begin{abstract}
Recent experiments on few-layer graphene structures have reported indirect signatures of anomalous Hall crystals (AHCs), but the need for a top gate to stabilize the phase precludes direct imaging of the emergent electronic lattice. This situation necessitates the investigation of alternative signatures of AHCs. The gapless phonons of the emergent electronic lattice provide a clear distinction from conventional quantum Hall states, but it may be difficult to disentangle these phonons from the plethora of other possible low-lying modes. Intriguingly, the quantum geometry of the underlying electronic ground state can imprint on the collective modes, possibly leading the phonons themselves to be topological. Were this the case, the resulting neutral chiral edge modes would provide a further signature of an AHC. Using time-dependent Hartree-Fock, we compute the spectra of collective modes of Wigner crystals (WCs) and AHCs arising in minimal models and study the topology of the phonons and low-lying excitons. Across the WC to AHC transition, we observe a series of band inversions among collective modes, producing topological phonons and excitons, and a sharp sign change in the phonon Chern number upon entering the AHC phase. We conclude by discussing the relevance of collective mode topology to experiments on candidate systems for AHCs.
\end{abstract}

\maketitle


Collective modes are a powerful probe of many-body systems, often serving as a revealing fingerprint of underlying broken symmetry or topological order that may otherwise be hard to access. In systems ranging from superfluids to quantum magnets and fractional quantum Hall liquids, the structure and dispersion of neutral excitations encode information about the nature of the system. The presence of gapless magnons, for example, indicates the breaking of continuous spin rotation symmetry~\cite{van1958spin, stancil2009spin}. The gapless mode of superfluids similarly marks the breaking of global $U(1)$ symmetry, while the massive roton mode contains critical information about the structure factor~\cite{bogoliubov1947theory, PhysRev.94.262, PhysRev.102.1189}. In the context of the quantum Hall effect, the softening of the magnetoroton mode indicates proximity to a phase transition, and the softening of the exotic graviton-like quadrupolar mode can precede a nematic ordering~\cite{girvinCollectiveExcitationGapFractional1985, girvinMagnetorotonTheoryCollective1986, PhysRevLett.107.116801, golkar2016spectral, regnaultEvidenceFractionalQuantum2017}.

The study of collective modes is especially important when direct probes of the ground state are not available. This is precisely the case for the anomalous Hall crystal (AHC), a phase that arises when strong interactions drive electrons to spontaneously crystallize into an insulator with a finite Chern number~\cite{PhysRevLett.132.236601, tanParentBerryCurvature2024, yuMoireFractionalChern2025, dong2024anomalous, dongStabilityAnomalousHall2024}.
Some evidence of AHCs has been reported in Bernal bilayer graphene (BBG)~\cite{seilerQuantumCascadeCorrelated2022c, seilerSignaturesSlidingWigner2025c}, and the extended integer quantum anomalous Hall (EIQAH) state in rhombohedral pentalayer graphene has been interpreted as potentially arising from a weakly pinned AHC~\cite{lu2024fractional, lu2025extended, dong2024anomalous, zhou2024fractional, dong2024theory, soejima2024anomalous, dongStabilityAnomalousHall2024, patri2024extended}. However, the presence of a top gate in these experiments precludes direct observation of the emergent electronic lattice, for example, via scanning tunneling microscopy.

Unlike conventional Chern insulators, AHCs host gapless phonons---Goldstone modes arising from the spontaneous breaking of continuous translation symmetry~\cite{doi:10.1073/pnas.2515532122, kwanMoireFractionalChern2025, tanIdealLimitRhombohedral2025, desrochers2025elasticresponseinstabilitiesanomalous}. Although observing electronic lattice phonons would offer a sharp signature of an AHC, it may be challenging to disentangle them from the myriad of other possible low-lying modes. It is natural to wonder if the topology and geometry of the underlying electronic ground state directly imprint on these collective modes. The Berry curvature of the phonons is finite in general, as time-reversal symmetry is spontaneously broken, leaving open the tantalizing possibility of the phonons acquiring a finite Chern number. This would offer a conceptually rich emergent realization of topological phonons, a phenomenon yet to be observed in two-dimensional atomic lattices~\cite{ zhangTopologicalNaturePhonon2010, qinBerryCurvaturePhonon2012a, jiTopologicalPhononModes2017}. If it were found that AHCs generically host topological electronic crystal phonons, the resulting neutral chiral edge modes spanning the gap between the phonons and gapped collective modes could serve as an alternative signature of an AHC.

In this work, we determine the topology of phonons and excitons of the AHC phases that arise in two toy models initially developed to study rhombohedral graphene. We numerically obtain the collective modes via time-dependent Hartree-Fock (TDHF) and compute the Chern numbers of the phonon and exciton modes directly from the TDHF wavefunctions. By varying the total Berry flux penetrating the first Brillouin zone (1BZ), we observe the collective mode topology evolve across the transition from Wigner crystal (WC), an electronic crystal with vanishing Chern number, to AHC. We also consider the impact that a weak periodic potential has on the topology of the collective modes, motivated by the ongoing debate regarding the significance of the moir\'e periodic potential that is present in recent experiments on rhombohedral graphene~\cite{dong2024theory, dong2024anomalous, zhou2024fractional, soejima2024anomalous, herzog-arbeitmanMoireFractionalChern2024, kwanMoireFractionalChern2025, yuMoireFractionalChern2025, li2025stackingorientationtwistanglecontrolinteger}. We conclude by discussing how electronic crystal phonon topology may be used to detect AHCs experimentally.

\begin{figure*}[t]
    \includegraphics[trim={0 6 0 7},clip]{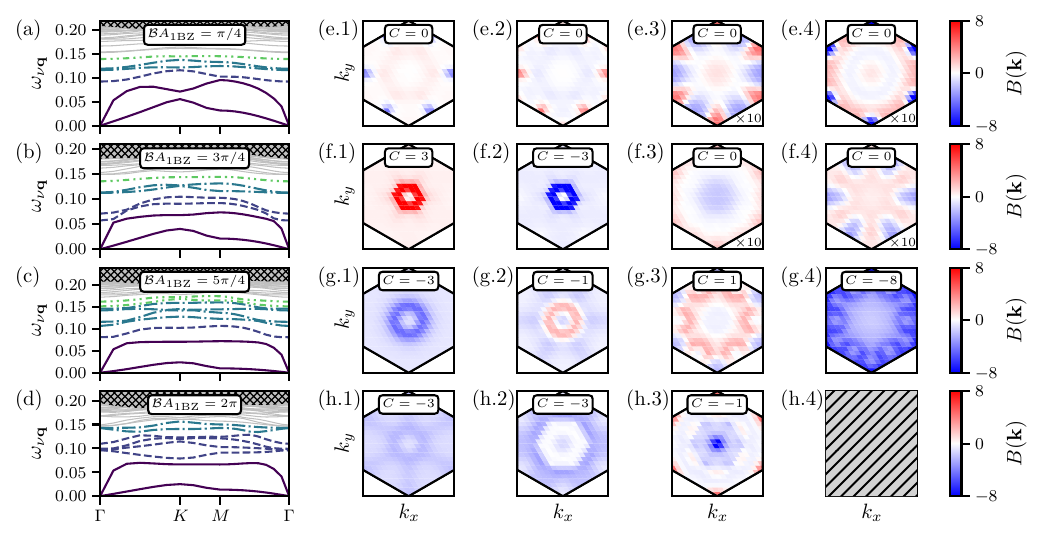}
    \caption{The collective mode spectrum along high-symmetry lines of the infinite Chern band model with $r_s=10$ and (a) $\mathcal{B}A_{\text{1BZ}}=\pi/4$, (b) $\mathcal{B}A_{\text{1BZ}}=3\pi/4$, (c) $\mathcal{B}A_{\text{1BZ}}=5\pi/4$, and (d) $\mathcal{B}A_{\text{1BZ}}=2\pi$. The lines indicate bands of collective modes, where line color and style are used to depict energetically isolated groups of bands, and the hatched region marks the non-interacting particle-hole continuum. The trace of the non-Abelian Berry curvature of the isolated groups of bands in (a-d) are plotted in (e-h), respectively. The panels of (e-h) are numbered in order of increasing energy of the bands, such that (x.1), (x.2), (x.3), and (x.4) correspond to the solid purple, dashed blue, dot-dashed turquoise, and dot-dot-dashed green lines, respectively. The Berry curvatures in (e.3-e.4) and (f.3-f.4) are scaled up by a factor of ten for visibility. The higher energy excitons of the AHC in (d) are too dense to resolve the Berry curvature, so we only plot the lowest three sets of bands in this case, leaving panel (h.4) blank.
    }
    \label{fig:figure_1}
\end{figure*}

\textit{Models and methods}.---
We consider continuum Hamiltonians of the form $\mathcal{H} = \mathcal{H}_0 + \mathcal{H}_{\text{int}}$, where $\mathcal{H}_0=\sum_{\vb{p}} \Cdag{\vb{p}} \mathcal{E}(\vb{p}) \C{\vb{p}}$ is the single-particle Hamiltonian and $\Cdag{\vb{p}}$ creates an electron with unbounded momentum $\vb{p}$. It is convenient to define the dimensionless Wigner-Seitz radius, $r_s$, the radius of a disk containing one electron, in units of the Bohr radius $a_B=4\pi\epsilon_0\hbar^2/e^2m$. We work in conventional density-dependent units, measuring distance in units of the average inter-electron distance $a_e=r_sa_B$ and setting $m=\hbar=e=1$, such that energies are in units of Rydbergs, $\text{Ry}=\frac{me^4}{2\hbar^2}$, and the Fermi momentum is $k_F=2$. The geometry of the band is encoded in the form factors $\mathcal{F}(\vb{p},\vb{p}') = \braket{s_{\vb{p}}}{s_{\vb{p}'}}$, where $\ket{s_{\vb{p}}} = e^{-i\vb{p}\cdot\vb{r}}\Cdag{\vb{p}}\ket{0}$ is the cell-periodic part of the Bloch wavefunction. We write the band-projected density-density interaction as $\mathcal{H}_{\text{int}} =\frac{1}{2A} \sum_{\vb{p}_1\vb{p}_2\vb{p}_3\vb{p}_4} \tilde{V}_{\vb{p}_1\vb{p}_2\vb{p}_3\vb{p}_4} \Cdag{\vb{p}_1} \Cdag{\vb{p}_2} \C{\vb{p}_3} \C{\vb{p}_4},$
where $A$ is the area of the system, $\tilde{V}_{\vb{p}_1\vb{p}_2\vb{p}_3\vb{p}_4}= V\left(\vb{p}_1 - \vb{p}_4\right) \mathcal{F}\left(\vb{p}_1, \vb{p}_4\right) \mathcal{F}\left(\vb{p}_2, \vb{p}_3 \right) \delta_{\vb{p}_1 + \vb{p}_2 - \vb{p}_3 - \vb{p}_4},$ and we take $V(\vb{p})=4\pi/r_sp$ as the unscreened Coulomb interaction.

We utilize two models developed for the study of AHCs, the infinite Chern band~\cite{tanParentBerryCurvature2024} and $\lambda-$jellium~\cite{soejimaJelliumModelAnomalous2025}. Both models have a simple quadratic dispersion, $\mathcal{E}(\vb{p})=p^2/r_s^2$, where $p=|\vb{p}|$, and both have ``ideal'' quantum geometry~\cite{PhysRevLett.127.246403, PhysRevB.108.205144}. The infinite Chern band has a constant Berry curvature $\Omega(\vb{p})=\mathcal{B}$ and possesses form factors identical to that of the lowest Landau level, $\mathcal{F}(\vb{p},\vb{p}')=\text{exp}\left[-\frac{\mathcal{B}}{4}\left(|\vb{p}-\vb{p}'|^2+2i\vb{p}\times\vb{p}'\right)\right]$, where $\vb{p}\times\vb{p}'\equiv p_xp_y'-p_yp_x'$. The Berry curvature of the $\lambda-$jellium model is given by $\Omega(\vb{p})= 2\lambda^2/\left(\lambda^2 p^2 + 1\right)^2$, where $\lambda$ tunes how sharply the Berry curvature is peaked at $p=0$. The corresponding form factor is $\mathcal{F}(\vb{p},\vb{p}')=f_{\vb{p}}f_{\vb{p}'}(1+2\lambda^2p_z p_{\bar{z}}')$, with $f_{\vb{p}}=1/\sqrt{1+\lambda^2 p^2}$ and $p_z=(p_x + ip_y)/\sqrt{2}$. In contrast to the unbounded Berry flux of the infinite Chern band, the $\lambda-$jellium model has a total Berry flux of $2\pi$ when integrated over the entire momentum space, regardless of the value of $\lambda$. Both models exhibit a phase transition from a Fermi liquid to an AHC, with HF predicting a critical value of $r_s\approx2$~\footnote{The critical value of $r_s$ is significantly underestimated by Hartree-Fock, but the finite Berry curvature of the parent band greatly improves the accuracy of the mean-field prediction compared to that of the Wigner crystal transition in a trivial two-dimensional electron gas~\cite{valentiQuantumGeometryDriven2025}.}. For the low interaction strengths considered here, the Chern number of the resulting AHC is well approximated by simple rounding of the Berry flux enclosed in the 1BZ to the nearest integer multiple of $2\pi$~\cite{dongStabilityAnomalousHall2024}.

We first solve these models via self-consistent Hartree-Fock (HF), imposing by hand the breaking of continuous to discrete translation symmetry. Anticipating the expense of subsequent TDHF calculations, we employ a modestly sized $18\times18$ triangular lattice and project into the lowest 19 bands before performing the HF~\footnote{The triangular lattice AHC is unstable in some cases, but is always the true ground state for the parameters considered here~\cite{kwanMoireFractionalChern2025, zhouNewClassesQuantum2025, desrochers2025elasticresponseinstabilitiesanomalous, doi:10.1073/pnas.2515532122}.}. Next, we employ the TDHF framework, constructing creation operators for the collective modes and solving for their energies~\cite{roweEquationsofMotionMethodExtended1968, khalafSoftModesMagic2020, doi:10.1073/pnas.2515532122}. In TDHF, the collective modes are assumed to be particle-hole excitations above some correlated ground state. We define the collective mode creation operator as $Q^\dagger_{\nu\vb{q}} = \sum_{\varphi, \vb{k}} \left(X^{\nu\vb{q}}_{\varphi}(\vb{k}) b^\dagger_{\varphi,\vb{q}}(\vb{k}) - Y^{\nu\vb{q}}_{\varphi}(\vb{k}) b^{\phantom{\dagger}}_{\varphi,-\vb{q}}(\vb{k})\right)$, where $\nu$ and $\vb{q}$ are the collective mode band index and momentum, $\varphi=(\phi_p, \phi_h)$ is a combined index denoting a pair of particle (unoccupied) and hole (occupied) HF bands, $b^\dagger_{\varphi,\vb{q}}(\vb{k}) = \eta^\dagger_{\phi_p}(\vb{k}+\vb{q})\eta_{\phi_h}(\vb{k})$ is an elementary particle-hole creation operator, $\eta^\dagger_{\phi}(\vb{k})$ creates an electron in HF band $\phi$ with momentum $\vb{k}$, $X^{\nu\vb{q}}_{\varphi}(\vb{k})$ and $Y^{\nu\vb{q}}_{\varphi}(\vb{k})$ are coefficients to be determined, and $\vb{k}$ and $\vb{q}$ are restricted to the 1BZ. When considering momenta outside the 1BZ, we make the definition $\eta^\dagger_\phi(\vb{k}) = \eta^\dagger_\phi(\lceil \vb{k} \rceil)$, where $\lceil \vb{k}\rceil$ folds the momentum back into the 1BZ. The correlated ground state, $\ket{0},$ is defined in TDHF as that which is annihilated by $Q_{\nu\vb{q}}$. The spectrum, along with $X^{\nu\vb{q}}_{\varphi}(\vb{k})$ and $Y^{\nu\vb{q}}_{\varphi}(\vb{k})$, are obtained by solving the equations of motion for $Q^\dagger_{\nu\vb{q}}$ in the quasi-boson approximation, i.e., assuming that the elementary particle-hole creation and annihilation operators obey bosonic commutation relations.

\begin{figure}[t]
\includegraphics[trim={0 8 0 7},clip]{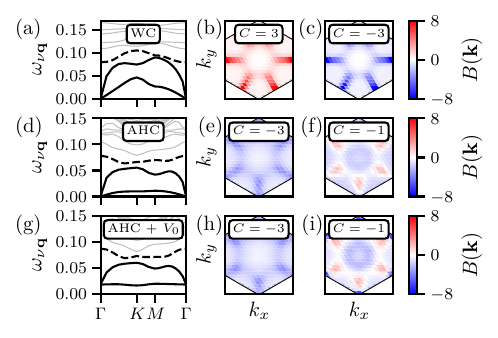}
\caption{
The collective mode spectrum along high-symmetry lines of the $\lambda-$jellium model for $r_s=10$ with (a) $\lambda=0.59$ (WC) and (d) $\lambda=2.65$ (AHC). The solid thick lines indicate the phonon modes, the dashed thick line is the lowest energy exciton, and the thin gray lines mark higher energy excitons. The trace of the non-Abelian Berry curvature of the phonons is plotted in (b, e) for $\lambda=0.59$ and $\lambda=2.65$, respectively, and the Berry curvature of the lowest energy excitons is similarly plotted in (c, f). The insets denote the Chern numbers of the bands. In panels (g-i) we plot the collective mode spectrum, phonon Berry curvature, and lowest exciton Berry curvature for the AHC with $\lambda=2.65$ including an external periodic potential of strength $V_0=2.5\times10^{-3}$.
}
\label{fig:figure_2}
\end{figure}

The correlated ground state consists of a coherent superposition of many particle-hole excitations above the HF ground state,
$\ket{0} = \text{exp}\left(\frac{1}{2}\sum_{\vb{q},\alpha,\beta}
    b^\dagger_{\vb{q},\alpha} Z^{\vb{q}}_{\alpha\beta} b^\dagger_{-\vb{q},\beta}\right)\ket{\text{HF}},$
where we again assume the quasi-boson approximation, $\alpha$ and $\beta$ are collective indices for $\vb{k}$ and $\varphi$, $\left[Z^{\vb{q}}\right]^\dagger = \vb{Y}^{\vb{q}}[\vb{X}^{\vb{q}}]^{-1}$, and $\vb{X}^{\vb{q}}$ is a matrix with components $\left[\vb{X}^{\vb{q}}\right]_{\alpha\nu} = X^{\nu\vb{q}}_\alpha$ (similar for $\vb{Y}^{\vb{q}}$)~\cite{gambacurtaExtensionSecondRandomphase2006}. Previous TDHF studies of collective mode Berry curvature have considered only gapped excitons, for which $Y^{\nu\vb{q}}$, and therefore $Z^{\vb{q}}$, is small~\cite{kwanExcitonBandTopology2021, PhysRevResearch.7.023047}. The collective mode wavefunctions, $\ket{\nu\vb{q}} \equiv Q_{\nu\vb{q}}^\dagger\ket{0},$ are well-approximated in this case by the form $\ket{\nu\vb{q}} \approx \mathcal{N}^\nu_{\vb{q}} \sum_{\varphi,\vb{k}}X^{\nu\vb{q}}_\varphi(\vb{k})b^\dagger_{\varphi,\vb{q}}(\vb{k})\ket{\text{HF}},$ where $\mathcal{N}^\nu_{\vb{q}}$ is a normalization constant. However, the phonon modes of electronic crystals have equal weight on $X^{\nu\vb{q}}$ and $Y^{\nu\vb{q}}$ at the gapless point $\vb{q}=0$, so this approximation cannot be applied. We instead expand $\ket{0}$ to first order in $Z^{\vb{q}}$, which gives collective mode wavefunctions $\ket{\nu\vb{q}} = \mathcal{N}^\nu_{\vb{q}} \sum_{\varphi,\vb{k}} \psi^{\nu,\vb{q}}_\varphi(\vb{k}) b^\dagger_{\varphi,\vb{q}}(\vb{k}) \ket{\text{HF}}$ with coefficients
\begin{equation}
        \psi^{\nu,\vb{q}}_\varphi(\vb{k})
        =
        X^{\nu\vb{q}}_\varphi(\vb{k}) - \sum_{\varphi',\vb{k}'}Z^{\vb{q}}_{\varphi\varphi'}(\vb{k},\vb{k}')Y^{\nu\vb{q}}_{\varphi'}(\vb{k}').
        \label{eq:approx_wf}
\end{equation}
Even with this modification of the phonon wavefunction, it is possible that the proliferation of particle-hole pairs caused by the finite value of $Z^{\vb{q}=0}$ spoils the validity of the correlated ground state itself. We confirm below that this is not the case by studying the collective mode topology in the presence of a weak periodic potential.

To compute the Chern number of collective modes, we must evaluate the Berry connection of the above eigenstates. Recent work revealed that there is an infinite family of possible Berry connections that can be defined for collective particle-hole excitations~\cite{kwanExcitonBandTopology2021, paivaShiftPolarizationExcitons2024, davenportBerryCurvatureLowEnergy2025, davenportExcitonBerryology2025}. The choice of Berry connection corresponds to a choice for the definition of the collective position coordinate of the excitation, parameterized by $\alpha\in[0, 1]$ as $\vb{R} = \alpha \vb{r}_h + (1-\alpha) \vb{r}_p,$ with $\vb{r}_{h/p}$ the position operators for the hole and particle, respectively. We choose $\alpha=1$ because we find it produces better numerical stability than $\alpha=0$, and avoid intermediate values of $\alpha$ as they require a momentum grid incommensurate with ours to compute the Berry curvature. We compute the non-Abelian Berry curvature by evaluating Wilson loops of the Berry connection around plaquettes in the 1BZ, and report the trace of the non-Abelian Berry curvature for isolated sets of bands, which enters in the calculation of the first Chern class~\footnote{See the Supplementary Information for more details.}.

\textit{Results}.---
We begin with the infinite Chern band, setting $r_s=10$, and consider both WC and AHC phases. For the WC, we study two values of the Berry flux, $\mathcal{B}A_{\text{1BZ}}=\pi/4$ and $3\pi/4,$ and for the AHC we take $\mathcal{B}A_{\text{1BZ}}=5\pi/4$ and $2\pi$, which produces an electronic Chern number $C_{\rm el}=1$.  We plot in Fig.~\ref{fig:figure_1}(a-d) the spectrum of collective modes for these four values of the Berry flux, finding that each case exhibits two gapless phonons and many exciton bands pulled down from the particle-hole continuum. In each band structure plot, we indicate with different line colors and styles the energetically isolated sets of bands for which we compute the trace of the non-Abelian Berry curvature and the Chern number. The Berry curvature for each set of bands is plotted in Fig. 1(e-h), in order of increasing energy from left to right, such that Fig.~\ref{fig:figure_1}(x.1) corresponds to the phonons, Fig.~\ref{fig:figure_1}(x.2) to the lowest energy isolated set of exciton bands, and so forth. The inset at the top of each panel denotes the Chern number of the corresponding set of bands.

For $\mathcal{B}A_{\text{1BZ}}=\pi/4$, the lowest two sets of bands exhibit equal and opposite peaks of Berry curvature at the $M$ point, where the upper phonon and lowest exciton band are closest in energy. However, because time-reversal symmetry has been only weakly broken, all of the bands are topologically trivial. Increasing the enclosed Berry flux to $\mathcal{B}A_{\text{1BZ}}=3\pi/4$, we see that the phonons and lowest set of excitons acquire equal and opposite Chern numbers, $C=\pm3$, respectively, resulting from the bands inverting at the $M$ point. The Berry curvature of these bands is again concentrated where the gap between them is smallest, on a ring around the $\Gamma$ point. In both cases, the Berry curvature of the upper exciton bands is very weak, so we scale it by a factor of ten for visibility.

Increasing the Berry flux further, there are no significant changes to the collective mode topology until the system transitions to the $C_{\rm el}=1$ AHC at  $\mathcal{B}A_{\text{1BZ}}=\pi$. Upon entering the phase, we find that the Chern number of the phonons abruptly changes sign to $C=-3$, as shown for $\mathcal{B}A_{\text{1BZ}}=5\pi/4$ in Fig.~\ref{fig:figure_1}(g.1). However, the transition from WC to AHC is first-order, so we cannot track the evolution of the band structure across it as we did within the WC phase. Probing deeper into the AHC phase, we see in Fig.~\ref{fig:figure_1}(h) that the Berry curvature of the phonons at $\mathcal{B}A_{\text{1BZ}}=2\pi$ becomes more uniform and the Chern number is still $C=-3$. As we show in the Supplementary Information, the phonon Chern number remains unchanged until $\mathcal{B}$ is tuned near to the boundary of the $C_{\rm el}=1$ AHC, where the upper phonon branch inverts with the lowest exciton at $\mathcal{B}A_{\text{1BZ}}\approx2.8\pi$, changing the phonon Chern number to $C=-1$~\cite{Note2}. For larger values of $\mathcal{B}$, the phonons become highly dispersive and the gap between them and the excitons is quite small, making it difficult to resolve the topology with currently accessible system sizes.

It is important to determine whether the physics observed in the infinite Chern band model persists in more realistic settings. The $\lambda-$jellium model provides a useful tool for this goal, as its finite Berry curvature distribution is more similar to that of relevant material platforms like BBG and R5G. We plot the in Fig.~\ref{fig:figure_2}(a-c) the collective mode spectrum, phonon Berry curvature, and lowest energy exciton Berry curvature of the WC that arises in this model at $r_s=10$ and $\lambda=0.59$. The physics observed here is qualitatively identical to that of the infinite Chern band WC: the upper phonon and lowest exciton bands have inverted at the $M$ point, yielding $C=\pm3$ for each band, respectively. Moving deep into the $C_e=1$ AHC phase at $\lambda=2.65$, we plot the collective mode spectrum and Berry curvature of the phonons and lowest exciton for $r_s=10$ and $\lambda=2.65$ in Fig.~\ref{fig:figure_2}(d-f), finding again that the Chern number of the phonons is the opposite sign as that of the WC. Although the low-energy excitons are well-separated, they still exhibit complicated Berry curvature distributions, and there is no obvious pattern to their Chern numbers.

Motivated by the lack of clarity on the role that the moir\'e potential plays in the EIQAH phase of R5G, we also compute the collective modes of AHCs in the presence of a weak periodic potential~\cite{dong2024theory, dong2024anomalous, zhou2024fractional, soejima2024anomalous, herzog-arbeitmanMoireFractionalChern2024, kwanMoireFractionalChern2025, yuMoireFractionalChern2025, li2025stackingorientationtwistanglecontrolinteger}. We plot in Fig.~\ref{fig:figure_2}(g-i) the spectrum and Berry curvature of the phonons and excitons of the $\lambda-$jellium AHC with $r_s=10$ and $\lambda=2.65$ in the presence of a periodic potential of strength $V_0=2.5\times10^{-3}$, taking the lattice to align with that of the spontaneously formed electronic lattice.

The periodic potential gaps out the phonons and splits their degeneracy at the $\Gamma$ point (the splitting is smaller than the line width), which dramatically reduces the value of $Z^{\vb{q}=0}$, placing the system in the regime where the approximation underlying Eq.~\eqref{eq:approx_wf} should hold. We observe that the presence of the periodic potential leaves the Berry curvature of the phonons and excitons nearly unchanged. It is reasonable to assume that the addition of a weak periodic potential doesn't affect the phonon topology. Given this, the above observation indicates that our approximation can correctly capture the topology of the phonons even in the gapless regime where $Z^{\vb{q}=0}$ may not be small. Further confirmation of this speculation could be achieved by studying the accuracy of the energetic obtained within this approximation, or by computing the Berry curvature of phonon wavefunctions accounting for higher powers of $Z^{\vb{q}}$ in the ground state. However, such calculations are beyond the scope of this work.

For both the AHC and the WC, we find that increasing the strength of the periodic potential always eventually leads to a band inversion between the phonons and excitons, leaving the phonons trivial. This trivialization of the phonons indicates that the periodic potential is playing a more significant role than simply pinning the spontaneously formed crystal. This observation also complements the understanding that the presence of Berry curvature in a parent band leads to drastic differences in the ground state topology resulting from spontaneous and explicit translation symmetry breaking~\cite{tanParentBerryCurvature2024, desrochers2025electroniccrystalphasespresence}

\textit{Discussion}.---
In this work, we showed that the quantum geometry of the parent band indeed imprints on the collective modes of electronic crystals, giving rise to topological phonons and excitons. We observe a suggestive pattern of changes in the phonon Chern number as we tune between WC and AHC phases, but ultimately find that the topology of the collective modes is not simply related to the topology of the ground state and is sensitive to, for example, crossings between phonon and exciton bands. We speculate that progress towards a predictive understanding of the relation between AHC ground state and phonon topology may be made by exploiting the recently identified mapping between AHCs in RNG and layer-pseudospin skyrmion lattices~\cite{tanIdealLimitRhombohedral2025}, as the topology of collective modes of skyrmion lattices is a well-studied topic~\cite{roldan-molinaTopologicalSpinWaves2016, diazTopologicalMagnonsEdge2019, diazChiralMagnonicEdge2020, weberTopologicalMagnonBand2022, akazawaTopologicalThermalHall2022, ghaderMomentumspaceTheoryTopological2024a}. We also observed that a strong periodic potential trivializes the gapped phonons of pinned AHCs, which is highly relevant to understanding the role that the moir\'e potential plays in the postulated AHC phase of R5G slightly misaligned on hBN. Specifically, if the presence of topological phonons in the absence of a moir\'e potential is theoretically established, but neutral chiral edge modes are not observed experimentally in the systems with a moir\'e potential, this would indicate that the hBN-induced moir\'e potential plays a more critical role than merely pinning and stabilizing the spontaneously formed electronic crystal.

\footnotetext[4]{We cannot confirm with currently accessible system sizes if this trend continues in the $C_e=3$ AHC phase of the infinite Chern band, and there is no AHC phase in the $\lambda-$jellium model with $C_e>1$.}

The question of how to experimentally detect electronic crystal phonon topology remains to be answered. In the absence of a top gate, scanning electron energy loss spectroscopy could detect the neutral chiral edge modes~\cite{songRecentProgressElectron2025}, but in this situation it is possible to directly image the electronic lattice via scanning tunneling microscopy. In principle, non-local heat transport mediated by neutral chiral edge modes could be measured via a scanning temperature probe~\cite{halbertalNanoscaleThermalImaging2016, doi:10.1126/science.aan0877} or a Hall bar of temperature probes~\cite{hirschbergerLargeThermalHall2015, banerjeeObservationHalfintegerThermal2018, czajkaPlanarThermalHall2023}, although these experiments are likely to be challenging. Finally, 
the finite Berry curvature of the phonons near the $\Gamma$ point can be detected through the temperature dependence of the anomalous thermal Hall effect at low temperatures~\cite{paivaShiftPolarizationExcitons2024}, although it must be disentangled from the $T-$linear integer quantum anomalous Hall contribution~\cite{PhysRevB.55.15832}.

Much important work remains to be done in studying the geometry and topology of collective modes in AHCs. Perhaps most urgent is extending these results to microscopic models of few-layer graphene structures. This presents a significant computational challenge, as obtaining the collective mode wavefunctions even for the single-band models considered here is quite expensive. It would also be of interest to compute other geometric quantities of collective modes in AHCs, such as the quantum geometric dipole moment~\cite{fertigManybodyQuantumGeometric2025, paivaShiftPolarizationExcitons2024, chenQuantumgeometricDipoleTopological2025}. Finally, the phonon topology could be informative for understanding how an AHC may melt into a proximate chiral superconductor, a scenario that may be relevant to multilayer rhombohedral graphene~\cite{han2025signatures, seo2025family, yang2025magnetic}.

\textit{Acknowledgments}.---
We thank Valentin Crépel, Daniel Parker, and Yves H. Kwan for insightful discussions. We acknowledge support from the Natural Sciences and Engineering Research Council of Canada (NSERC) and the Centre for Quantum Materials at the University of Toronto. Computations were performed on the Cedar and Fir clusters, which the Digital Research Alliance of Canada hosts.


%

\end{document}


\title{Supplementary Information for ``Topological phonons in anomalous Hall crystals''}

\author{Mark R. Hirsbrunner \orcidlink{0000-0001-8115-6098}}
\email{mark.hirsbrunner@utoronto.ca}
\affiliation{
Department of Physics, University of Toronto, Toronto, Ontario M5S 1A7, Canada
}

\author{F\'elix Desrochers \orcidlink{0000-0003-1211-901X}}
\affiliation{
Department of Physics, University of Toronto, Toronto, Ontario M5S 1A7, Canada
}
\affiliation{
Department of Physics, Harvard University, Cambridge, MA 02138, United States}

\author{Joe Huxford \orcidlink{0000-0002-4857-0091}}
\affiliation{
Department of Physics, University of Toronto, Toronto, Ontario M5S 1A7, Canada
}
\affiliation{
Department of Physics and Astronomy, University of Manchester, Oxford Road, Manchester M13 9PL, United Kingdom
}

\author{Yong Baek Kim}
\email{yongbaek.kim@utoronto.ca}
\affiliation{
 Department of Physics, University of Toronto, Toronto, Ontario M5S 1A7, Canada
}

\date{\today}

\maketitle

\renewcommand{\figurename}{Supplementary Figure}
\renewcommand{\thefigure}{\arabic{figure}}
\counterwithin*{figure}{part}
\stepcounter{part}

\section{Large-$\mathcal{B}$ phonon topology}
In Supplementary Fig.~\ref{fig:figure_S1}, we plot the collective mode spectrum and phonon Berry curvature for the $C_e=1$ AHC of the infinite Chern band at $\mathcal{B}A_{\text{1BZ}}=2.75\pi$ and $\mathcal{B}A_{\text{1BZ}}=2.9\pi$. The Chern number of the phonons changes from $C=3$ to $C=-1$ between these two values of $\mathcal{B}$, driven by a series of band inversions between the excitons and phonons.

\begin{figure}
\includegraphics[]{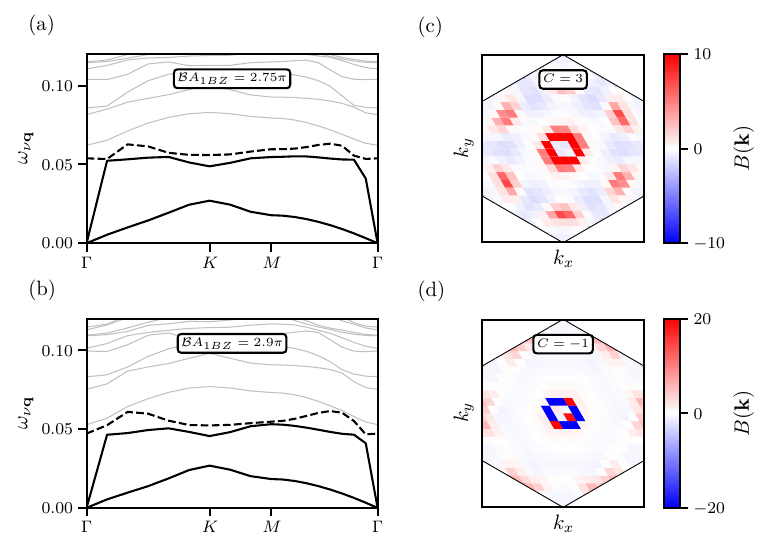}
\caption{
The collective mode spectrum along high-symmetry lines of the $C_e=1$ AHC arising in the infinite Chern band for $r_s=10$ with (a) $\mathcal{B}A_{\text{1BZ}}=2.75\pi$ and (b) $\mathcal{B}A_{\text{1BZ}}=2.9\pi$. The solid thick lines indicate the phonon modes, the dashed thick line is the lowest energy exciton, and the thin gray lines mark higher energy excitons. The Berry curvature of the phonons is plotted in (c, d) for $\mathcal{B}A_{\text{1BZ}}=2.75\pi$ and $\mathcal{B}A_{\text{1BZ}}=2.9\pi$, respectively. The insets denote the Chern numbers of the bands.
}
\label{fig:figure_S1}
\end{figure}

\section{Hartree-Fock}
Anticipating the crystallization transition, let us first write the Hamiltonian, $\mathcal{H} = \mathcal{H}_0 + \mathcal{H}_{\text{int}}$, in terms of crystal momentum and reciprocal lattice vectors. The single-particle Hamiltonian takes the form
\begin{equation}
    \mathcal{H}_0 = \sum_{\vb{k},\vb{g}} \Cdag{\vb{k}, \vb{g}} \mathcal{E}(\vb{k}+\vb{g}) \C{\vb{k},\vb{g}}
    -
    \sum_{\vb{k}}\sum_{\vb{g},\vb{g}'} c^\dagger_{\vb{k},\vb{g}+\vb{g}'} V_{\vb{g}'} \mathcal{F}(\vb{k}+\vb{g}+\vb{g}', \vb{k}+\vb{g}) \C{\vb{k},\vb{g}},
\end{equation}
where we have included a $C_6-$symmetric periodic potential, keeping only the first harmonics, i.e. $V_{\vb{g}}=V_0\delta_{\vb{g},\vb{G}_i}$ with $\vb{G}_i$ denoting the six smallest reciprocal lattice vectors (excluding $\vb{g}=\vb{0}$). It can be turned on to model a moir\'e lattice aligned with the spontaneously formed lattice. We assume a triangular lattice and take the primitive reciprocal lattice vectors to be $\vb{g}_1 = g(1, 0)$ and $\vb{g}_2 = g(-\frac{1}{2}, \frac{\sqrt{3}}{2}),$ with $g=4\pi/a \sqrt{3}$ and lattice constant $a^2=2\pi/\sqrt{3}.$ We include all momentum $|\vb{g}| \leq 5g$ when summing over reciprocal lattice vectors.
The interaction term is given by
\begin{equation}
    \mathcal{H}_{\text{int}}
    =
    \frac{1}{2A}\sum_{\substack{\vb{k}_1\vb{k}_2\vb{k}_3\vb{k}_4 \\ \vb{g}_1\vb{g}_2\vb{g}_3\vb{g}_4}}V^{\vb{g}_1\vb{g}_2\vb{g}_3\vb{g}_4}_{\vb{k}_1\vb{k}_2\vb{k}_3\vb{k}_4} \Cdag{\vb{k}_1,\vb{g}_1}\Cdag{\vb{k}_2,\vb{g}_2}\C{\vb{k}_3,\vb{g}_3}\C{\vb{k}_4,\vb{g}_4},
\end{equation}
with coefficients
\begin{equation}
    \begin{aligned}
        V^{\vb{g}_1\vb{g}_2\vb{g}_3\vb{g}_4}_{\vb{k}_1\vb{k}_2\vb{k}_3\vb{k}_4} &= V\left(\vb{k}_1 + \vb{g}_1 - \vb{k}_4 - \vb{g}_4 \right) \mathcal{F}\left(\vb{k}_1+\vb{g}_1, \vb{k}_4 + \vb{g}_4\right) \mathcal{F}\left(\vb{k}_2 + \vb{g}_2, \vb{k}_3 + \vb{g}_3 \right)\delta_{\vb{k}_1 + \vb{g}_1 + \vb{k}_2 + \vb{g}_2 - \vb{k}_3 - \vb{g}_3 -\vb{k}_4 - \vb{g}_4}.
    \end{aligned}
\end{equation}
Before performing the mean-field decoupling, we diagonalize the single-particle Hamiltonian (trivial without the inclusion of a periodic potential),
\begin{equation}
    \begin{aligned}
        \mathcal{H}_0 &= \sum_{\vb{k}} \sum_{\vb{g},\vb{g}'} H_{\vb{g}\vb{g'}}(\vb{k})\Cdag{\vb{k},\vb{g}}\C{\vb{k},\vb{g}'}= \sum_{\vb{k}, a}\mathcal{E}_a(\vb{k})d^\dagger_{\vb{k},a}d_{\vb{k},a}.
    \end{aligned}
\end{equation}
Here $a$ labels the single-particle bands, and the Hamiltonian is diagonalized by the operators
\begin{equation}
    \begin{gathered}
        d^\dagger_{\vb{k},a} = \sum_{\vb{g}}U_{\vb{g}a}^{\text{sp}}(\vb{k}) \Cdag{\vb{k},\vb{g}}, \quad \Cdag{\vb{k}, \vb{g}} = \sum_a U_{a\vb{g}}^{\text{sp}\dagger} d^\dagger_{\vb{k},a}
        \\
        d_{\vb{k},a} = \sum_{\vb{g}}U_{a\vb{g}}^{\text{sp}\dagger}(\vb{k}) \C{\vb{k},\vb{g}}, \quad \C{\vb{k}, \vb{g}} = \sum_a U_{\vb{g}a}^{\text{sp}} d_{\vb{k},a}.
    \end{gathered}
\end{equation}
To reduce the computational expense of subsequent tdHF calculations, we project the entire Hamiltonian into the lowest $n_b=19$ bands of the single-particle Hamiltonian, corresponding to a cutoff of two momentum shells. The projected Hamiltonian is then
\begin{equation}
    \mathcal{H} = \sum_{\vb{k}}\sum_{a=1}^{n_b}\mathcal{E}_a(\vb{k})d^\dagger_{\vb{k},a}d_{\vb{k},a}
    +
    \frac{1}{2A}\sum_{\vb{k}_1\vb{k}_2\vb{k}_3\vb{k}_4}\sum_{a,b,c,d=1}^{n_b}V_{\vb{k}_1\vb{k}_2\vb{k}_3\vb{k}_4}^{abcd} d^\dagger_{\vb{k}_1,a} d^\dagger_{\vb{k}_2,b} d_{\vb{k}_3,c} d_{\vb{k}_4,d},
\end{equation}
where the matrix elements of the Coulomb interaction are
\begin{equation}
    V_{\vb{k}_1\vb{k}_2\vb{k}_3\vb{k}_4}^{abcd}
    =
    \sum_{\vb{g}_1\vb{g}_2\vb{g}_3\vb{g}_4}
    U_{a\vb{g}_1}^{\text{sp}\dagger}(\vb{k}_1) U_{b\vb{g}_2}^{\text{sp}\dagger}(\vb{k}_2)
    V^{\vb{g}_1\vb{g}_2\vb{g}_3\vb{g}_4}_{\vb{k}_1\vb{k}_2\vb{k}_3\vb{k}_4}
    U_{\vb{g}_3c}^{\text{sp}}(\vb{k}_3) U_{\vb{g}_4d}^{\text{sp}}(\vb{k}_4)
\end{equation}

Next we perform a mean-field decoupling of the Coulomb interaction, yielding the Hartree and Fock terms
\begin{equation}
    H_{\text{H}} = \frac{1}{A}\sum_{\vb{k},\vb{k}'} \sum_{a,b,c,d=1}^{n_b}
    V^{abcd}_{\vb{k}\vb{k}'\vb{k}'\vb{k}}
    \mathcal{P}_{ad}(\vb{k})
    d^\dagger_{\vb{k}',b}d^{\phantom{\dagger}}_{\vb{k}',c}
\end{equation}
\begin{equation}
    H_{\text{F}} = -\frac{1}{A}\sum_{\vb{k},\vb{k}'}\sum_{a,b,c,d=1}^{n_b}
    V^{abcd}_{\vb{k}\vb{k}'\vb{k}\vb{k}'}
    \mathcal{P}_{ac}(\vb{k})
    d^\dagger_{\vb{k}',b}d^{\phantom{\dagger}}_{\vb{k}',d},
\end{equation}
where $\mathcal{P}_{ab}(\vb{k})=\expval{d^\dagger_{\vb{k},a}d_{\vb{k},b}}$ is the density matrix.The density matrix is in one-to-one correspondence with Slater determinants, so HF corresponds to optimizing the energy over the space of Slater determinants. We accomplish this optimization by solving the HF Hamiltonian, $H_{\text{HF}} = H_0 + H_{\text{H}} + H_{\text{F}},$ self-consistently for the density matrix. We start with a randomly initialized density matrix and employ periodic Pulay mixing for the optimization~\cite{pulay1980convergence, pulay1982improved, kresse1996efficient, rohwedder2011analysis}.

The resulting Hamiltonian is diagonalized by the transformation
\begin{equation}
    \begin{gathered}
        \eta^\dagger_{\phi}(\vb{k}) = \sum_{b}U^{\text{HF}}_{b\phi}(\vb{k})d^\dagger_{\vb{k},b}, \,\quad d^\dagger_{\vb{k},a} = \sum_{\phi}U^{\text{HF}\dagger}_{\phi a}(\vb{k})\eta^\dagger_\phi(\vb{k}),
        \\
        \eta_{\phi}(\vb{k}) = \sum_{b}U^{\text{HF}\dagger}_{\phi b}(\vb{k})d_{\vb{k},b} ,\,\quad d_{\vb{k},a} = \sum_{\phi}U^{\text{HF}}_{a \phi}(\vb{k})\eta_\phi(\vb{k}),
    \end{gathered}
\end{equation}
where $\phi$ labels the Hartree-Fock bands and columns of $U^{\text{HF}}_{a\phi}(\vb{k})$ are the eigenvectors of the HF Hamiltonian. The above transformations can be combined into a single unitary, $U_{\vb{g}\phi}(\vb{k}) = \sum_a U^{\text{sp}}_{\vb{g} a}(\vb{k}) U^{\text{HF}}_{a\phi}(\vb{k})$, directly relating the HF bands to the Bloch waves. When considering momenta outside the first Brillouin zone, we make the definitions
\begin{equation}
    \begin{gathered}
        \eta_{\phi}(\vb{k}) = \eta_{\phi}(\lceil\vb{k}\rceil),
        \\
        U_{\vb{g}\phi}(\vb{k} + \vb{g}') = U_{\vb{g} + \vb{g}',\phi}(\vb{k}),
    \end{gathered}
\end{equation}
where $\lceil\vb{k}\rceil$ folds the momentum back to the first Brillouin zone.

The Hartree-Fock energies are given by
\begin{equation}
    \mathcal{E}^{\text{HF}}_{\phi}(\vb{k})\delta_{\phi'\phi}
    =
    \tilde{\mathcal{E}}_{\phi'\phi}(\vb{k}) + \frac{1}{A} \sum_{\vb{p}, \phi_h} 
    \left(
    \tilde{V}^{\phi'\phi_h\phi_h\phi}_{\vb{k}\vb{p}\vb{p}\vb{k}}
        -
    \tilde{V}^{\phi_h\phi'\phi_h\phi}_{\vb{p}\vb{k}\vb{p}\vb{k}}
    \right)
    ,
\end{equation}
where we have defined
\begin{equation}
    \begin{aligned}
        \tilde{\mathcal{E}}_{\phi \phi'}(\vb{k})
        &\equiv
        \sum_{a=1}^{n_b}U^{\text{HF}\dagger}_{\phi a}(\vb{k})\mathcal{E}_{a}(\vb{k})U^{\text{HF}}_{a\phi'}(\vb{k})
    \end{aligned}
\end{equation}
and
\begin{equation}
    \begin{aligned}
        \tilde{V}^{\phi\phi'\varphi\varphi'}_{\vb{k}_1\vb{k}_2\vb{k}_3\vb{k}_4}
        &\equiv
        \sum_{a,b,c,d=1}^{n_b} U^{\text{HF}\dagger}_{\phi a}(\vb{k}_1) U^{\text{HF}\dagger}_{\phi' b}(\vb{k}_2) V^{abcd}_{\vb{k}_1\vb{k}_2\vb{k}_3\vb{k}_4} U^{\text{HF}}_{c\varphi}(\vb{k}_3) U^{\text{HF}}_{d\varphi'}(\vb{k}_4).
        \\
        &=\sum_{\vb{g}_1\vb{g}_2\vb{g}_3\vb{g}_4} 
        U^{\dagger}_{\phi \vb{g}_1}(\vb{k}_1) U^{\dagger}_{\phi' \vb{g}_2}(\vb{k}_2)
        V^{\vb{g}_1\vb{g}_2\vb{g}_3\vb{g}_4}_{\vb{k}_1\vb{k}_2\vb{k}_3\vb{k}_4}
        U_{\vb{g}_3\varphi}(\vb{k}_3)
        U_{\vb{g}_4\varphi'}(\vb{k}_4).
    \end{aligned}
\end{equation}
The HF ground state is obtained by filling the $N$ lowest energy orbitals. In the basis of the HF bands, the full Hamiltonian takes the form
\begin{equation}
    \mathcal{H} = \sum_{\vb{k}} \sum_{\phi,\phi'=1}^{n_b} \tilde{\mathcal{E}}_{\phi\phi'}(\vb{k})\eta^\dagger_{\phi}(\vb{k})\eta_{\phi'}(\vb{k})
    +
    \frac{1}{2A}\sum_{\vb{k}_1\vb{k}_2\vb{k}_3\vb{k}_4} \sum_{\phi, \phi'=1}^{n_b} \sum_{\vartheta, \vartheta'=1}^{n_b}\tilde{V}^{\phi \phi' \vartheta \vartheta'}_{\vb{k}_1\vb{k}_2\vb{k}_3\vb{k}_4} \eta^\dagger_{\phi}(\vb{k}_1)\eta^\dagger_{\phi'}(\vb{k}_2)\eta_{\vartheta}(\vb{k}_3)\eta_{\vartheta'}(\vb{k}_4).
\end{equation}

\section{Time-dependent Hartree-Fock}
In TDHF, we seek to obtain collective excitations above the ground state in the form
\begin{equation}
    Q^\dagger_{\nu\vb{q}} = \sum_{\varphi, \vb{k}} \left(X^{\nu\vb{q}}_{\varphi}(\vb{k}) b^\dagger_{\varphi,\vb{q}}(\vb{k}) - Y^{\nu\vb{q}}_{\varphi}(\vb{k}) b^{\phantom{\dagger}}_{\varphi,-\vb{q}}(\vb{k})\right).
\end{equation}
where $\varphi$ labels pairs of particle and hole bands, $\varphi=(\phi_p, \phi_h)$, and $b^\dagger_{\varphi,\vb{q}}(\vb{k})=\eta^\dagger_{\phi_p}(\vb{k}+\vb{q})\eta_{\phi_h}(\vb{k})$ is the creation operator for an elementary particle-hole pair with momentum $\vb{q}.$ We assume these excitations to be eigenstates of the many-body Hamiltonian, obeying the Schr\"odinger equation $\left[\mathcal{H}, Q^\dagger_{\nu\vb{q}}\right] = \hbar\omega_{\nu\vb{q}}Q^\dagger_{\nu\vb{q}}$. Starting from the Schr\"odinger equations, it can be shown through elementary manipulations that the following relation holds for any operator $R$,
\begin{equation}
    \mel{0}{[R,[\mathcal{H},Q^\dagger_{\nu\vb{q}}]]}{0} = \hbar\omega_{\nu\vb{q}}\mel{0}{[R,Q^\dagger_{\nu\vb{q}}]}{0}.
\end{equation}
where $\ket{0}$ is the correlated ground state, defined as that which is annihilated by all $\hat{Q}_{\nu\vb{q}}.$ Taking $R$ to be elementary particle-hole creation and annihilation operators, one obtains the equations of motion for the collective excitations,
\begin{equation}
    \sum_{\vb{k},\varphi}
    \begin{pmatrix}
        A^{\vb{q}}_{(\varphi',\varphi)}(\vb{k}',\vb{k}) & B^{\vb{q}}_{(\varphi',\varphi)}(\vb{k}',\vb{k}) \\
        B^{-\vb{q}*}_{(\varphi',\varphi)}(\vb{k}',\vb{k}) & A^{-\vb{q}*}_{(\varphi',\varphi)}(\vb{k}',\vb{k})
    \end{pmatrix}
    \begin{pmatrix}
        X^{\nu\vb{q}}_{\varphi}(\vb{k}) \\
        Y^{\nu\vb{q}}_{\varphi}(\vb{k})
    \end{pmatrix}
    =
    \omega_{\nu\vb{q}}
    \begin{pmatrix}
        1 & 0 \\
        0 & -1
    \end{pmatrix}
    \begin{pmatrix}
        X^{\nu\vb{q}}_{\varphi'}(\vb{k}') \\
        Y^{\nu\vb{q}}_{\varphi'}(\vb{k}')
    \end{pmatrix}
    \label{eq:eom}
\end{equation}
where the matrices $A^{\vb{q}}$ and $B^{\vb{q}}$ take the form
\begin{equation}
    A_{(\varphi',\varphi)}^{\vb{q}}(\vb{k}', \vb{k}) = \mel{0}{[b^{\phantom{\dagger}}_{\varphi',\vb{q}}(\vb{k}'), [H, b^\dagger_{\varphi,\vb{q}}(\vb{k})]]}{0}
\end{equation}
and
\begin{equation}
    B_{(\varphi',\varphi)}^{\vb{q}}(\vb{k}',\vb{k}) = -\mel{0}{[b^{\phantom{\dagger}}_{\varphi',\vb{q}}(\vb{k}'), [H, b^{\phantom{\dagger}}_{\varphi,-\vb{q}}(\vb{k})]]}{0}.
\end{equation}

To obtain an explicit, numerically tractable form of the equations of motion, we take the quasi-boson approximation, assuming that the particle-hole operators obey bosonic commutation relations, i.e.
\begin{equation*}
    \comm{b^{\phantom{\dagger}}_{\varphi,\vb{q}}(\vb{k})}{b^\dagger_{\varphi',\vb{q}'}(\vb{k}')}\ket{0}
    \approx
    \comm{b^{\phantom{\dagger}}_{\varphi,\vb{q}}(\vb{k})}{b^\dagger_{\varphi',\vb{q}'}(\vb{k}')}\ket{\text{HF}}
    = \delta_{\varphi\varphi'}\delta_{\vb{q}\vb{q}'}\delta_{\vb{k}\vb{k}'} \ket{\text{HF}}.
\end{equation*}
This amounts to replacing the correlated ground state $\ket{0}$ with the HF ground state $\ket{\text{HF}}$, which produces the following expressions for $A^{\vb{q}}$ and $B^{\vb{q}}$,
\begin{equation}
    \begin{aligned}
        A^{\vb{q}}_{(\varphi',\varphi)}(\vb{k}',\vb{k})
        &=
        \frac{1}{A} \tilde{V}^{\phi_h\phi_p'\phi_h'\phi_p}_{\vb{k}\lceil\vb{k}'+\vb{q}\rceil\vb{k}'\lceil\vb{k}+\vb{q}\rceil}
        -
        \frac{1}{A} \tilde{V}^{\phi_p'\phi_h\phi_h'\phi_p}_{\lceil\vb{k}'+\vb{q}\rceil\vb{k}\vb{k}'\lceil\vb{k}+\vb{q}\rceil}
        +
        \left(\mathcal{E}^{\text{HF}}_{\phi_p}(\lceil\vb{k}+\vb{q}\rceil) - \mathcal{E}^{\text{HF}}_{\phi_h}(\vb{k})\right)\delta_{\vb{k}', \vb{k}}\delta_{\phi_p'\phi_p}\delta_{\phi_h'\phi_h}
    \end{aligned}
\end{equation}
and
\begin{equation}
    B^{\vb{q}}_{(\varphi',\varphi)}(\vb{k}',\vb{k})
    =
    \frac{1}{A}\left( \tilde{V}^{\phi_p'\phi_p\phi_h\phi_h'}_{\lceil\vb{k}'+\vb{q}\rceil\lceil\vb{k}-\vb{q}\rceil\vb{k}\vb{k}'}
    -
    \tilde{V}^{\phi_p'\phi_p\phi_h'\phi_h}_{\lceil\vb{k}'+\vb{q}\rceil\lceil\vb{k}-\vb{q}\rceil\vb{k}'\vb{k}} \right).
\end{equation}
These expressions can be explicitly evaluated, and eigenvalues and eigenvectors of the equations of motion can be obtained by numerically solving Eq.~\eqref{eq:eom}.

\section{Approximate collective mode wavefunctions}
The standard approach for computing the wavefunctions of excitons in TDHF is the straightforward application of the quasi-boson approximation, i.e., replacing the correlated ground state with the HF ground state in the definition of the wavefunction: $\ket{\nu,\vb{q}}= \hat{Q}^\dagger_{\nu\vb{q}}\ket{0} \approx \hat{Q}^\dagger_{\nu\vb{q}}\ket{\text{HF}}.$ As mentioned in the main text, this approach fails for phonons because it disregards the hole-particle component of the mode, captured by the vector $Y^{\nu\vb{q}}(\vb{k})$, which annihilates the HF ground state. We improve upon this situation by accounting to first order for the coherent superposition of particle and hole states that form the correlated ground state.

The correlated ground state in the quasi-boson approximation takes the form
\begin{equation}
    \ket{0} = \text{exp}\left(\frac{1}{2}\sum_{\varphi,\varphi'}\sum_{\vb{k},\vb{k}'}\sum_{\vb{q}}
    Z^{\vb{q}}_{\varphi'\varphi}(\vb{k}',\vb{k})b^\dagger_{\varphi',\vb{q}}(\vb{k}')b^\dagger_{\varphi,-\vb{q}}(\vb{k})\right)\ket{\text{HF}},
\end{equation}
where the matrix $Z^{\vb{q}}$ is given by
\begin{equation}
    Z^{\vb{q}} = \left(\vb{Y}^{\vb{q}}[\vb{X}^{\vb{q}}]^{-1}\right)^\dagger
\end{equation}
and $\vb{X}^{\vb{q}} = [X^{0\vb{q}}, X^{1\vb{q}}, \dots, X^{N\vb{q}}]$, i.e. $\vb{X}^{\vb{q}}$ is a matrix with elements $\left[\vb{X}^{\vb{q}}\right]_{\alpha\nu} = X^{\nu\vb{q}}_\alpha$ (similar for $\vb{Y}^{\vb{q}}$)~\cite{gambacurtaExtensionSecondRandomphase2006}. To account for the presence of particle-hole pairs in the ground state, we expand the exponential to first order in $Z^{\vb{q}}$,
\begin{equation}
    \ket{0} \approx \left[1 + \frac{1}{2}\sum_{\varphi,\varphi'}\sum_{\vb{k},\vb{k}'}\sum_{\vb{q}}
    Z^{\vb{q}}_{\varphi'\varphi}(\vb{k}',\vb{k})b^\dagger_{\varphi',\vb{q}}(\vb{k}')b^\dagger_{\varphi,-\vb{q}}(\vb{k})\right]\ket{\text{HF}}.
\end{equation}
Acting on this approximate ground state with the collective mode creation operator gives the form of the collective mode wavefunctions that we employ to compute the Berry curvature,
\begin{equation}
    \begin{aligned}
        \ket{\nu\vb{q}}
        &\approx
        \mathcal{N}^\nu_{\vb{q}} \sum_{\varphi,\vb{k}}\left(X^{\nu\vb{q}}_\varphi(\vb{k}) - \sum_{\varphi',\vb{k}'}Z^{\vb{q}}_{\varphi\varphi'}(\vb{k},\vb{k}')Y^{\nu\vb{q}}_{\varphi'}(\vb{k}')\right)b^\dagger_{\varphi,\vb{q}}(\vb{k})\ket{\text{HF}}
        \\
        &=
        \mathcal{N}^\nu_{\vb{q}} \sum_{\varphi,\vb{k}} \psi^{\nu,\vb{q}}_\varphi(\vb{k}) b^\dagger_{\varphi,\vb{q}}(\vb{k}) \ket{\text{HF}}.
    \end{aligned}
\end{equation}

Because $Z^{\vb{q}}$ is not small at the gapless point of the phonon modes, this is not necessarily an accurate approximation of the phonon wavefunctions. The singular values of $Z^{\vb{q}}$, which control the relative weights of the zero and two particle-hole pair sectors of the ground state, can be made small with the addition of a periodic potential that gaps the phonons. We show in the main text that the Berry curvature remains qualitatively unchanged with the addition of such a periodic potential, indicating that the jump from including zero particle-hole pairs in the ground state to including up to two does not impact the Berry curvature. It remains to be seen how the inclusion of higher numbers of particle-hole pairs in the ground state changes the Berry curvature, but we leave this to future work, as constructing the Berry curvature for such many-body states is currently an open question. Finally, we note that our primary interest is in the topology of the phonons, which is robust to errors in the Berry curvature at an isolated point.

\section{Collective mode Berry curvature}
In Refs.~\cite{davenportExcitonBerryology2025, davenportBerryCurvatureLowEnergy2025} the Berry connection and curvature of excitons in two-band systems is derived by constructing exciton Wannier functions that maximally localize either the particle- or hole-projected position operator. The derivation proceeds by first building a periodic position operator for either the particle or hole bands and projecting it into the exciton bands. The authors then show that the eigenvectors of these projected position operators are exciton Wannier functions that maximally localize either the particle or the hole in the exciton. The eigenvalues of the periodic position operators are shown to take the form of excitonic generalizations of electronic Wilson loops, from which the Berry connection can be obtained by taking the thermodynamic limit. The exciton Berry curvature can be obtained by evaluating these Wilson loops around plaquettes in the Brillouin zone~\cite{fukui2005chern}. In this section we generalize this approach to obtain the non-Abelian Berry connection and Berry curvature for collective modes of multi-band systems.

The density operator is defined as
\begin{equation*}
    \begin{aligned}
        \hat{n}(\vb{R} + \vb{r})
        &=
        \sum_{\vb{k},\vb{k}'} \sum_{\vb{g},\vb{g}'}
        e^{-i(\vb{k}+\vb{g})\cdot(\vb{R}+\vb{r})}
        \mathcal{F}(\vb{k}+\vb{k}'+\vb{g}+\vb{g}',\vb{k}'+\vb{g}')
        c^\dagger_{\vb{k}+\vb{k}',\vb{g}+\vb{g}'}c^{\phantom{\dagger}}_{\vb{k}',\vb{g}'}
        \\
        &=
        \sum_{\vb{k},\vb{k}'} \sum_{\vb{g},\vb{g}'} \sum_{\phi,\phi'}
        e^{-i(\vb{k}+\vb{g})\cdot(\vb{R}+\vb{r})}
        U^\dagger_{\phi,\vb{g}+\vb{g}'}(\vb{k}+\vb{k}')
        \mathcal{F}(\vb{k}+\vb{k}'+\vb{g}+\vb{g}',\vb{k}'+\vb{g}')
        U_{\vb{g}'\phi'}(\vb{k}')
        \eta^\dagger_{\phi}(\vb{k}+\vb{k}') \eta^{\phantom{\dagger}}_{\phi'}(\vb{k}'),
    \end{aligned}
\end{equation*}
where $\vb{R}$ denotes a lattice site and $\vb{r}$ the position within the unit cell. The band projected particle density operator is obtained by restricting the sum over the bands to only the particle bands,
\begin{equation}
    \hat{n}^p_{\vb{R}+\vb{r}}
    =
    \sum_{\vb{k},\vb{k}'} \sum_{\vb{g},\vb{g}'} \sum_{\phi_p,\phi_p'}
    e^{-i(\vb{k}+\vb{g})\cdot(\vb{R}+\vb{r})}
    U^\dagger_{\phi_p,\vb{g}+\vb{g}'}(\vb{k}+\vb{k}')
    \mathcal{F}(\vb{k}+\vb{k}'+\vb{g}+\vb{g}',\vb{k}'+\vb{g}')
    U_{\vb{g}'\phi_p'}(\vb{k}')
    \eta^\dagger_{\phi_p}(\vb{k}+\vb{k}') \eta^{\phantom{\dagger}}_{\phi_p'}(\vb{k}'),
\end{equation}
and the band-projected hole density operator is obtained by swapping the order of the creation and annihilation operators and restricting the sum to only the hole bands,
\begin{equation}
    \hat{n}^h_{\vb{R}+\vb{r}}
    =
    \sum_{\vb{k},\vb{k}'} \sum_{\vb{g},\vb{g}'} \sum_{\phi_h,\phi_h'}
    e^{-i(\vb{k}+\vb{g})\cdot(\vb{R}+\vb{r})}
    U^\dagger_{\phi_h,\vb{g}+\vb{g}'}(\vb{k}+\vb{k}')
    \mathcal{F}(\vb{k}+\vb{k}'+\vb{g}+\vb{g}',\vb{k}'+\vb{g}')
    U_{\vb{g}'\phi_h'}(\vb{k}')
    \eta^{\phantom{\dagger}}_{\phi_h'}(\vb{k}') \eta^\dagger_{\phi_h}(\vb{k}+\vb{k}').
\end{equation}
Using these, we can construct periodic position operators as
\begin{equation}
    \hat{Z}_{p/h} = \frac{1}{L^2}\sum_{\vb{R}} \int_{\text{u.c.}}d^2\vb{r}\,  e^{i\vb{\delta}\cdot(\vb{R}+\vb{r})} \hat{n}^{p/h}_{\vb{R}+\vb{r}},
\end{equation}
where $\vb{\delta}$ points along a primitive reciprocal lattice vector and has magnitude $\delta = |\vb{\delta}|= 2\pi/L$. Here $L=Na$ is the linear size of the system, with $a$ the lattice constant and $N$ the number of sites along this direction. To proceed, we further project $\hat{Z}_{p/h}$ into some energetically isolated set of collective modes,
\begin{equation}
    \hat{\mathcal{Z}}_{p/h} = \hat{P}_{\text{col}}\hat{Z}_{p/h}\hat{P}_{\text{col}},
\end{equation}
where
\begin{equation}
    \hat{P}_{\text{col}} = \sum_{\mu,\vb{q}}\hat{Q}^\dagger_{\mu\vb{q}}\ketbra{0}{0}\hat{Q}^{\phantom{\dagger}}_{\mu\vb{q}}
\end{equation}
and the sum over $\mu$ is restricted to the desired modes.

To obtain the collective mode Wilson loop and, subsequently, the Berry connection, we need to compute the eigenvalues of this periodic position operator. Focusing first on the particle-projected position operator, we have
\begin{equation}
    \begin{aligned}
        \left[\hat{\mathcal{Z}}_p\right]^{\mu\nu}_{\vb{q},\vb{q}'}
        &=
        \frac{1}{L^2} \bar{\mathcal{N}}^\mu_{\vb{q}} \mathcal{N}^\nu_{\vb{q}'}
        \sum_{\vb{k},\vb{k}'} \sum_{\vb{\ell},\vb{\ell}'} \sum_{\vb{g},\vb{g}'} \sum_{\varphi,\varphi'} \sum_{\phi_p'',\phi_p'''}
        \sum_{\vb{R}}\int_{\text{u.c.}}d^2\vb{r}\,
        e^{-i(\vb{\ell}+\vb{g}-\vb{\delta})\cdot(\vb{R}+\vb{r})}
        \bar{\psi}^{\mu,\vb{q}}_\varphi(\vb{k}) \psi^{\nu, \vb{q}'}_{\varphi'}(\vb{k}')
        \\
        &\quad\times
        U^\dagger_{\phi_p'',\vb{g}+\vb{g}'}(\vb{\ell}+\vb{\ell}')
        \mathcal{F}(\vb{\ell}+\vb{\ell}'+\vb{g}+\vb{g}',\vb{\ell}'+\vb{g}')
        U_{\vb{g}'\phi_p'''}(\vb{\ell}')
        \\
        &\quad\times
        \mel{\text{HF}}{\eta^\dagger_{\phi_h}(\vb{k}) \eta^{\phantom{\dagger}}_{\phi_p}(\vb{k}+\vb{q}) \eta^\dagger_{\phi_p''}(\vb{\ell}+\vb{\ell}') \eta^{\phantom{\dagger}}_{\phi_p'''}(\vb{\ell}') \eta^\dagger_{\phi_p'}(\vb{k}'+\vb{q}') \eta^{\phantom{\dagger}}_{\phi_h'}(\vb{k}')}{\text{HF}}
        \\
        &=
        \frac{1}{L^2}\bar{\mathcal{N}}^\mu_{\vb{q}} \mathcal{N}^\nu_{\vb{q}'}
        \sum_{\vb{k},\vb{k}'} \sum_{\vb{\ell},\vb{\ell}'} \sum_{\vb{g},\vb{g}'} \sum_{\varphi,\varphi'} \sum_{\phi_p'',\phi_p'''}
        \sum_{\vb{R}}\int_{\text{u.c.}}d^2\vb{r}\,
        e^{-i(\vb{\ell}+\vb{g}-\vb{\delta})\cdot(\vb{R}+\vb{r})}
        \bar{\psi}^{\mu,\vb{q}}_\varphi(\vb{k}) \psi^{\nu, \vb{q}'}_{\varphi'}(\vb{k}')
        \\
        &\quad\times
        U^\dagger_{\phi_p'',\vb{g}+\vb{g}'}(\vb{\ell}+\vb{\ell}')
        \mathcal{F}(\vb{\ell}+\vb{\ell}'+\vb{g}+\vb{g}',\vb{\ell}'+\vb{g}')
        U_{\vb{g}'\phi_p'''}(\vb{\ell}')
        \\
        &\quad\times
        \delta_{\vb{k},\vb{k}'} \delta_{\vb{\ell}',\lceil \vb{k}'+\vb{q}' \rceil} \delta_{\lceil \vb{\ell}+\vb{\ell}' \rceil,\lceil \vb{k}+\vb{q} \rceil }\delta_{\phi_h,\phi_h'}\delta_{\phi_p,\phi_p''}\delta_{\phi_p',\phi_p'''}
        \\
        &=
        \frac{1}{L^2}\bar{\mathcal{N}}^\mu_{\vb{q}} \mathcal{N}^\nu_{\vb{q}'}
        \sum_{\vb{k}} \sum_{\vb{\ell},\vb{\ell}'} \sum_{\vb{g},\vb{g}'} \sum_{\phi_p,\phi_p'} \sum_{\phi_h}
        \sum_{\vb{R}}\int_{\text{u.c.}}d^2\vb{r}\,
        e^{-i(\lceil \vb{k}+\vb{q} \rceil - \lceil \vb{k}+\vb{q}' \rceil +\vb{g}_{\vb{\ell}+\vb{\ell}'} + \vb{g}-\vb{\delta})\cdot(\vb{R}+\vb{r})}
        \bar{\psi}^{\mu,\vb{q}}_{\phi_p, \phi_h} \psi^{\nu, \vb{q}'}_{\phi_p', \phi_h}(\vb{k})
        \\
        &\quad\times
        U^\dagger_{\phi_p,\vb{g}+\vb{g}'}(\lceil \vb{k}+\vb{q} \rceil + \vb{g}_{\vb{\ell}+\vb{\ell}'})
        \mathcal{F}(\lceil \vb{k}+\vb{q} \rceil + \vb{g}_{\vb{\ell}+\vb{\ell}'}+\vb{g}+\vb{g}',\lceil \vb{k}+\vb{q}' \rceil+\vb{g}')
        U_{\vb{g}'\phi_p'}(\lceil \vb{k}+\vb{q}' \rceil)
        \\
        &\quad\times
        \delta_{\vb{\ell}',\lceil \vb{k}'+\vb{q}' \rceil} \delta_{\lceil \vb{\ell}+\vb{\ell}' \rceil,\lceil \vb{k}+\vb{q} \rceil }
        \\
        &=
        \frac{1}{L^2}\bar{\mathcal{N}}^\mu_{\vb{q}} \mathcal{N}^\nu_{\vb{q}'}
        \sum_{\vb{k}} \sum_{\vb{\ell},\vb{\ell}'} \sum_{\vb{g},\vb{g}'} \sum_{\phi_p,\phi_p'} \sum_{\phi_h}
        \sum_{\vb{R}}\int_{\text{u.c.}}d^2\vb{r}\,
        e^{-i(\lceil \vb{k}+\vb{q} \rceil - \lceil \vb{k}+\vb{q}' \rceil +\vb{g}_{\vb{\ell}+\vb{\ell}'} + \vb{g}-\vb{\delta})\cdot(\vb{R}+\vb{r})}
        \bar{\psi}^{\mu,\vb{q}}_{\phi_p, \phi_h} \psi^{\nu, \vb{q}'}_{\phi_p', \phi_h}(\vb{k})
        \\
        &\quad\times
        U^\dagger_{\phi_p,\vb{g}+\vb{g}' + \vb{g}_{\vb{\ell}+\vb{\ell}'}}(\lceil \vb{k}+\vb{q} \rceil)
        \mathcal{F}(\lceil \vb{k}+\vb{q} \rceil + \vb{g}_{\vb{\ell}+\vb{\ell}'}+\vb{g}+\vb{g}',\lceil \vb{k}+\vb{q}' \rceil+\vb{g}')
        U_{\vb{g}'\phi_p'}(\lceil \vb{k}+\vb{q}' \rceil)
        \\
        &\quad\times
        \delta_{\vb{\ell}',\lceil \vb{k}'+\vb{q}' \rceil} \delta_{\lceil \vb{\ell}+\vb{\ell}' \rceil,\lceil \vb{k}+\vb{q} \rceil }
        \\
        &=
        \frac{1}{L^2}
        \bar{\mathcal{N}}^\mu_{\vb{q}} \mathcal{N}^\nu_{\vb{q}'}
        \sum_{\vb{k}} \sum_{\vb{g},\vb{g}'} \sum_{\phi_p,\phi_p'} \sum_{\phi_h}
        \sum_{\vb{R}}\int_{\text{u.c.}}d^2\vb{r}\,
        e^{-i(\lceil \vb{k} + \vb{q} \rceil - \lceil \vb{k} + \vb{q}' \rceil + \vb{g} - \vb{\delta})\cdot(\vb{R}+\vb{r})}
        \bar{\psi}^{\mu,\vb{q}}_{\phi_p, \phi_h}(\vb{k}) \psi^{\nu, \vb{q}'}_{\phi_p', \phi_h}(\vb{k})
        \\
        &\quad\times
        U^\dagger_{\phi_p,\vb{g}+\vb{g}'}(\lceil \vb{k} + \vb{q} \rceil)
        \mathcal{F}(\lceil \vb{k} + \vb{q} \rceil + \vb{g} + \vb{g}',\lceil \vb{k}+ \vb{q}' \rceil+\vb{g}')
        U_{\vb{g}'\phi_p'}(\lceil \vb{k} + \vb{q}' \rceil)
        \\
        &=
        \frac{N_{\text{u.c.}}}{L^2}\delta_{\vb{q}, \vb{q}' + \vb{\delta}}
        \bar{\mathcal{N}}^\mu_{\vb{q}' + \vb{\delta}} \mathcal{N}^\nu_{\vb{q}'}
        \sum_{\vb{k}} \sum_{\vb{g},\vb{g}'} \sum_{\phi_p,\phi_p'} \sum_{\phi_h}
        \int_{\text{u.c.}}d^2\vb{r}\,
        e^{-i(\vb{g} - \vb{g}_{\vb{k}+\vb{q}' +\vb{\delta}} + \vb{g}_{\vb{k}+\vb{q}'})\cdot \vb{r}}
        \bar{\psi}^{\mu,\vb{q}' + \vb{\delta}}_{\phi_p, \phi_h}(\vb{k}) \psi^{\nu, \vb{q}'}_{\phi_p', \phi_h}(\vb{k})
        \\
        &\quad\times
        U^\dagger_{\phi_p,\vb{g}+\vb{g}'}(\lceil \vb{k} + \vb{q}' + \vb{\delta} \rceil)
        \mathcal{F}(\lceil \vb{k} + \vb{q}' + \vb{\delta} \rceil + \vb{g} + \vb{g}',\lceil \vb{k}+ \vb{q}' \rceil+\vb{g}')
        U_{\vb{g}'\phi_p'}(\lceil \vb{k} + \vb{q}' \rceil)
        \\
        &=
        \delta_{\vb{q}, \vb{q}' + \vb{\delta}}
        \bar{\mathcal{N}}^\mu_{\vb{q}' + \vb{\delta}} \mathcal{N}^\nu_{\vb{q}'}
        \sum_{\vb{k}} \sum_{\vb{g}} \sum_{\phi_p,\phi_p'} \sum_{\phi_h}
        \bar{\psi}^{\mu,\vb{q}' + \vb{\delta}}_{\phi_p, \phi_h}(\vb{k}) \psi^{\nu, \vb{q}'}_{\phi_p', \phi_h}(\vb{k})
        \\
        &\quad\times
        U^\dagger_{\phi_p,\vb{g}+\vb{g}_{\vb{k}+\vb{q}'+\vb{\delta}}}(\lceil \vb{k} + \vb{q}' + \vb{\delta} \rceil)
        \mathcal{F}(\lceil \vb{k} + \vb{q}' + \vb{\delta} \rceil + \vb{g} + \vb{g}_{\vb{k}+\vb{q}'+\vb{\delta}},\lceil \vb{k}+ \vb{q}' \rceil+\vb{g} + \vb{g}_{\vb{k}+\vb{q}'})
        U_{\vb{g} + \vb{g}_{\vb{k}+\vb{q}'}, \phi_p'}(\lceil \vb{k} + \vb{q}' \rceil)
        \\
        &=
        \delta_{\vb{q}, \vb{q}' + \vb{\delta}}
        \bar{\mathcal{N}}^\mu_{\vb{q}' + \vb{\delta}} \mathcal{N}^\nu_{\vb{q}'}
        \sum_{\vb{k}} \sum_{\phi_p,\phi_p'} \sum_{\phi_h}
        \bar{\psi}^{\mu,\vb{q}' + \vb{\delta}}_{\phi_p, \phi_h}(\vb{k}) \psi^{\nu, \vb{q}'}_{\phi_p', \phi_h}(\vb{k})
        \\
        &\quad\times
        \sum_{\vb{g}} U^\dagger_{\phi_p\vb{g}}(\vb{k} + \vb{q}' + \vb{\delta})
        \mathcal{F}(\vb{k} + \vb{q}' + \vb{\delta} + \vb{g},\vb{k}+ \vb{q}' + \vb{g})
        U_{\vb{g}\phi_p'}(\vb{k} + \vb{q}')
        \\
        &=
        \delta_{\vb{q},\vb{q}'+\vb{\delta}}
        \bar{\mathcal{N}}^\mu_{\vb{q}} \mathcal{N}^\nu_{\vb{q}'}
        \sum_{\vb{k}} \sum_{\phi_h} \sum_{\phi_p,\phi_p'}
        \bar{\psi}^{\mu,\vb{q}'+\vb{\delta}}_\varphi(\vb{k}-\vb{q}') \psi^{\nu, \vb{q}'}_{\varphi'}(\vb{k}-\vb{q}')
        \braket{u_{\phi_p}(\vb{k}+\vb{\delta})}{u_{\phi_p'}(\vb{k})},
    \end{aligned}
\end{equation}
where we made use of the overlap between Bloch eigenstates, $\braket{u_{\phi}(\vb{k})}{u_{\phi'}(\vb{k}')} = \mel{\text{vac}}{\eta^{\phantom{\dagger}}_{\phi}(\vb{k}) e^{i(\vb{k} - \vb{k}')\cdot\vb{r}} \eta^\dagger_{\phi'}(\vb{k}')}{\text{vac}}$ with $\ket{\text{vac}}$ the vacuum state, and $N_{\text{u.c.}}$ is the number of unit cells in the system. Making the definition
\begin{equation}
    t_{p, \vb{q}}^{\mu\nu} = \bar{\mathcal{N}}^\mu_{\vb{q}+\vb{\delta}} \mathcal{N}^\nu_{\vb{q}}
        \sum_{\vb{k}} \sum_{\phi_h} \sum_{\phi_p,\phi_p'} 
        \bar{\psi}^{\mu,\vb{q}+\vb{\delta}}_{\phi_h, \phi_p}(\vb{k}-\vb{q}) \psi^{\nu, \vb{q}}_{\phi_h, \phi_p'}(\vb{k}-\vb{q})
        \braket{u_{\phi_p}(\vb{k}+\vb{\delta})}{u_{\phi_p'}(\vb{k})},
\end{equation}
we can write the projected periodic position operator concisely as
\begin{equation}
    \hat{\mathcal{Z}}=\sum_{\vb{q}} \sum_{\mu,\nu} t_{p,\vb{q}}^{\mu\nu} \ketbra{\mu \vb{q}+\vb{\delta}}{\nu\vb{q}}.
\end{equation}

The eigenvectors of the periodic position operator take the form
\begin{equation}
    \ket{\mathcal{W}_{p}^{\vb{R}}} = \sum_{\nu,\vb{q}} \mathcal{W}^{\vb{R},\nu}_{p}(\vb{q})\ket{\nu\vb{q}},
\end{equation}
and the periodic position operator acts on it as
\begin{equation*}
    \begin{aligned}
        \hat{\mathcal{Z}}\ket{\mathcal{W}_{p}^{\vb{R}}}
        &=
        \sum_{\nu,\vb{q}} \mathcal{W}^{\vb{R},\nu}_{p}(\vb{q}) \hat{\mathcal{Z}}\ket{\nu\vb{q}}
        \\
        &=
        \sum_{\mu, \nu, \vb{q}} \mathcal{W}^{\vb{R},\nu}_{p}(\vb{q}) t^{\mu\nu}_{p, \vb{q}} \ket{\mu\vb{q}+\vb{\delta}}
        \\
        &=
        \sum_{\mu, \nu, \vb{q}} \mathcal{W}^{\vb{R},\nu}_{p}(\vb{q}-\vb{\delta}) t^{\mu\nu}_{p, \vb{q}-\vb{\delta}} \ket{\mu\vb{q}}
        \\
        &=
        \lambda_{\vb{R}} \sum_{\nu,\vb{q}} \mathcal{W}^{\vb{R},\nu}_{p}(\vb{q}) \ket{\nu\vb{q}},
    \end{aligned}
\end{equation*}
where $\lambda_{\vb{R}}$ is the eigenvalue. For the last equality to hold, we require
\begin{equation}
    \sum_{\mu} t^{\nu\mu}_{p, \vb{q}} \mathcal{W}^{\vb{R},\mu}_{p}(\vb{q}) 
    =
    \lambda_{\vb{R}} \mathcal{W}^{\vb{R},\nu}_{p}(\vb{q} + \vb{\delta}),
\end{equation}
and similarly we have
\begin{equation}
    \sum_{\mu} t^{\nu\mu}_{p, \vb{q}+\vb{\delta}} \mathcal{W}^{\vb{R},\mu}_{p}(\vb{q}+\vb{\delta}) 
    =
    \lambda_{\vb{R}} \mathcal{W}^{\vb{R},\nu}_{p}(\vb{q} + 2\vb{\delta}).
\end{equation}
Substituting one into the other produces
\begin{equation}
    \mathcal{W}^{\vb{R},\nu}_{p}(\vb{q} + 2\vb{\delta})
    =
    \frac{1}{\lambda_{\vb{R}}^2}\sum_{\mu_1,\mu_2} t^{\nu\mu_2}_{p, \vb{q}+\vb{\delta}} t^{\mu_2\mu_1}_{p, \vb{q}} \mathcal{W}^{\vb{R},\mu_1}_{p}(\vb{q}),
\end{equation}
and repeating this process $N$ times yields
\begin{equation}
    \begin{aligned}
        \mathcal{W}^{\vb{R},\nu}_{p}(\vb{q} + 2\pi\hat{\delta})
        &=
        \frac{1}{\lambda_{\vb{R}}^N}\sum_{\mu_1,\dots,\mu_N} t^{\nu\mu_N}_{p, \vb{q}+2\pi\hat{\delta} - \vb{\delta}} t^{\mu_N\mu_{N-1}}_{p, \vb{q} + 2\pi\hat{\delta} - 2\vb{\delta}} \dots t^{\mu_3\mu_2}_{p, \vb{q} + \vb{\delta}} t^{\mu_2\mu_1}_{p, \vb{q}} \mathcal{W}^{\vb{R},\mu_1}_{p}(\vb{q})
        \\
        &=
        \frac{1}{\lambda_{\vb{R}}^N}\sum_{\mu} \mathcal{W}^{\nu\mu}_{p}(\vb{q}) \mathcal{W}^{\vb{R},\mu}_{p}(\vb{q}),
    \end{aligned}
\end{equation}
where $\hat{\delta} = \vb{\delta}/\delta$ and we have defined the exciton Wilson loop operator
\begin{equation}
    \mathcal{W}^{\nu\mu}_{p}(\vb{q}) = \sum_{\sigma_1,\dots,\sigma_{N-1}} t^{\nu\sigma_{N-1}}_{p, \vb{q}+2\pi\hat{\delta} - \vb{\delta}} t^{\sigma_{N-1}\sigma_{N-2}}_{p, \vb{q} + 2\pi\hat{\delta} - 2\vb{\delta}} \dots t^{\sigma_2\sigma_1}_{p, \vb{q} + \vb{\delta}} t^{\sigma_1\mu}_{p, \vb{q}}.
\end{equation}
Making use of the periodicity of $\vb{q}$, we have the new eigenvalue equation
\begin{equation}
    \sum_{\nu} \mathcal{W}^{\mu\nu}_p(\vb{q}) \mathcal{W}^{\vb{R},\nu}_{p}(\vb{q})
    =
    \lambda_{\vb{R}}^N \mathcal{W}^{\vb{R},\mu}_{p}(\vb{q}),
\end{equation}
which states that the eigenvalues of the periodic position operator are the $N^{\text{th}}$ roots of the eigenvalues of the exciton Wilson loop operator.

By analogy to the single-particle electronic case, we can obtain the exciton Berry curvature by taking the thermodynamic limit of the exciton Wilson loop. Taking $\vb{\delta}$ to be small, we have
\begin{equation}
    \begin{aligned}
        t_p^{\mu\nu}(\vb{q})
        &=
        \sum_{\vb{k}} \sum_{\phi_h} \sum_{\phi_p,\phi_p'} 
        \bar{\mathcal{N}}^\mu_{\vb{q}+\vb{\delta}}\bar{\psi}^{\mu,\vb{q}+\vb{\delta}}_{\phi_h, \phi_p}(\vb{k}-\vb{q}) \mathcal{N}^\nu_{\vb{q}}\psi^{\nu, \vb{q}}_{\phi_h, \phi_p'}(\vb{k}-\vb{q})
        \braket{u_{\phi_p}(\vb{k}+\vb{\delta})}{u_{\phi_p'}(\vb{k})}
        \\
        &=
        \sum_{\vb{k}} \sum_{\phi_h} \sum_{\phi_p,\phi_p'} 
        \bar{\mathcal{N}}^\mu_{\vb{q}+\vb{\delta}}\bar{\psi}^{\mu,\vb{q}+\vb{\delta}}_{\phi_h, \phi_p}(\vb{k}) \mathcal{N}^\nu_{\vb{q}}\psi^{\nu, \vb{q}}_{\phi_h, \phi_p'}(\vb{k})
        \braket{u_{\phi_p}(\vb{k}+\vb{q}+\vb{\delta})}{u_{\phi_p'}(\vb{k}+\vb{q})}
        \\
        &\approx
        \sum_{\vb{k}} \sum_{\phi_h} \sum_{\phi_p,\phi_p'} 
        \left(
        \bar{\mathcal{N}}^\mu_{\vb{q}} \bar{\psi}^{\mu,\vb{q}}_{\phi_h, \phi_p}(\vb{k})
        +
        \vb{\delta}\cdot\vb{\nabla}_{\vb{q}} \bar{\mathcal{N}}^\mu_{\vb{q}}\bar{\psi}^{\mu,\vb{q}}_{\phi_h, \phi_p}(\vb{k})
        \right)
        \mathcal{N}^\nu_{\vb{q}}\psi^{\nu, \vb{q}}_{\phi_h, \phi_p'}(\vb{k})
        \\
        &\times
        \left(
        \braket{u_{\phi_p}(\vb{k}+\vb{q})}{u_{\phi_p'}(\vb{k}+\vb{q})}
        +
        \braket{\vb{\delta}\cdot\vb{\nabla}_{\vb{k}} u_{\phi_p}(\vb{k}+\vb{q})}{u_{\phi_p'}(\vb{k}+\vb{q})}
        \right)
        \\
        &=
        \sum_{\vb{k}} \sum_{\phi_h} \sum_{\phi_p,\phi_p'} 
        \left(
        \bar{\mathcal{N}}^\mu_{\vb{q}} \bar{\psi}^{\mu,\vb{q}}_{\phi_h, \phi_p}(\vb{k})
        +
        \vb{\delta}\cdot\vb{\nabla}_{\vb{q}} \bar{\mathcal{N}}^\mu_{\vb{q}} \bar{\psi}^{\mu,\vb{q}}_{\phi_h, \phi_p}(\vb{k})
        \right)
        \mathcal{N}^\nu_{\vb{q}} \psi^{\nu, \vb{q}}_{\phi_h, \phi_p'}(\vb{k})
        \\
        &\times
        \left(
        \delta_{\phi_p,\phi_p'}
        -
        \braket{u_{\phi_p}(\vb{k}+\vb{q})}{\vb{\delta}\cdot\vb{\nabla}_{\vb{k}} u_{\phi_p'}(\vb{k}+\vb{q})}
        \right)
        \\
        &=
        \sum_{\vb{k}} \sum_{\phi_h} \sum_{\phi_p,\phi_p'}
        \bar{\mathcal{N}}^\mu_{\vb{q}}\bar{\psi}^{\mu,\vb{q}}_{\phi_h, \phi_p}(\vb{k})
        \left(
        \delta_{\phi_p,\phi_p'}(1 -\vb{\delta}\cdot\vb{\nabla}_{\vb{q}}) - \braket{u_{\phi_p}(\vb{k}+\vb{q})}{\vb{\delta}\cdot\vb{\nabla}_{\vb{k}} u_{\phi_p'}(\vb{k}+\vb{q})}
        \right)
        \mathcal{N}^\nu_{\vb{q}}\psi^{\nu, \vb{q}}_{\phi_h, \phi_p'}(\vb{k})
        \\
        &=
        \delta^{\mu\nu}
        +
        i\vb{\delta}\cdot\sum_{\phi_h} \sum_{\phi_p,\phi_p'}\int\frac{d^2\vb{k}}{\delta^2}
        \bar{\mathcal{N}}^\mu_{\vb{q}}\bar{\psi}^{\mu,\vb{q}}_{\phi_h, \phi_p}(\vb{k})
        \left(
        i\delta_{\phi_p,\phi_p'}\vb{\nabla}_{\vb{q}}
        -
        \vb{A}_{\phi_p\phi_p'}(\vb{k}+\vb{q})
        \right)
        \mathcal{N}^\nu_{\vb{q}}\psi^{\nu, \vb{q}}_{\phi_h, \phi_p'}(\vb{k})
        \\
        &=
        \left[\text{exp}\left(
        i\vb{\delta}\cdot\vb{\mathcal{A}}_p(\vb{q})\right)\right]^{\mu\nu},
    \end{aligned}
\end{equation}
where the electronic Berry connection is given by $\vb{A}_{\phi'\phi}(\vb{k}) = -i\braket{u_{\phi'}(\vb{k})}{\vb{\nabla}_{\vb{k}}u_{\phi}(\vb{k})}$ and we define the exciton Berry connection as
\begin{equation}
    \vb{\mathcal{A}}_p(\vb{q})^{\mu\nu}
    =
    \sum_{\phi_h} \sum_{\phi_p,\phi_p'}\int\frac{d^2\vb{k}}{\delta^2}
        \bar{\mathcal{N}}^\mu_{\vb{q}}\bar{\psi}^{\mu,\vb{q}}_{\phi_h, \phi_p}(\vb{k})
        \left(
        i\delta_{\phi_p,\phi_p'}\vb{\nabla}_{\vb{q}}
        -
        \vb{A}_{\phi_p\phi_p'}(\vb{k}+\vb{q})
        \right)
        \mathcal{N}^\nu_{\vb{q}}\psi^{\nu, \vb{q}}_{\phi_h, \phi_p'}(\vb{k}).
\end{equation}

Repeating the above derivation starting with the hole-projected position operator similarly yields a similar Wilson loop, defined as $\mathcal{W}^{\nu\mu}_{h}(\vb{q}) = \sum_{\sigma_1,\dots,\sigma_{L-1}} t^{\nu\sigma_{L-1}}_{h, \vb{q}+2\pi\hat{\delta} - \vb{\delta}} t^{\sigma_{L-1}\sigma_{L-2}}_{h, \vb{q} + 2\pi\hat{\delta} - 2\vb{\delta}} \dots t^{\sigma_2\sigma_1}_{h, \vb{q} + \vb{\delta}} t^{\sigma_1\mu}_{h, \vb{q}},$ with coefficients
\begin{equation}
    t_h^{\mu\nu}(\vb{q})
    =
    \bar{\mathcal{N}}^\mu_{\vb{q}+\vb{\delta}} \mathcal{N}^\nu_{\vb{q}}
        \sum_{\vb{k}} \sum_{\phi_p} \sum_{\phi_h,\phi_h'}
        \bar{\psi}^{\mu,\vb{q}+\vb{\delta}}_{\phi_h,\phi_p}(\vb{k}) \psi^{\nu, \vb{q}}_{\phi_h',\phi_p}(\vb{k}+\vb{\delta})
        \braket{u_{\phi_h'}(\vb{k}+\vb{\delta})}{u_{\phi_h}(\vb{k})}.
\end{equation}
The corresponding exciton Berry connection takes the form
\begin{equation}
    \vb{\mathcal{A}}^{\mu\nu}_h(\vb{q})
    =
    \bar{\mathcal{N}}^\mu_{\vb{q}}\mathcal{N}^\nu_{\vb{q}}
    \sum_{\phi_p} \sum_{\phi_h,\phi_h'}\int\frac{d^2\vb{k}}{\delta^2}
    \bar{\psi}^{\mu, \vb{q}}_{\phi_h, \phi_p}(\vb{k}-\vb{q})
    \left(
    i\delta_{\phi_h,\phi_h'}\vb{\nabla}_{\vb{q}}
    -
    \vb{A}_{\phi_h\phi_h'}(\vb{k}-\vb{q})
    \right)
    \psi^{\nu, \vb{q}}_{\phi_h', \phi_p}(\vb{k}-\vb{q}).
\end{equation}
As stated in the main text, $\vb{\mathcal{A}}_p(\vb{q})$ is associated with exciton Wannier functions that maximally localize the particle position operator, while $\vb{\mathcal{A}}_h(\vb{q})$ is associated with those that maximally localize the hole position operator.

To obtain the Berry curvature, we integrate the Wilson loop around plaquettes and take its determinant~\cite{fukui2005chern, barry_notes}. To do so in a numerically stable way, we make the definitions
\begin{equation}
    t_{p,a}^{\mu\nu}(\vb{q})
    =
    \sum_{\phi_h} \sum_{\phi_p,\phi_p'} \sum_{\vb{k}} \bar{\psi}_{\phi_p,\phi_h}^{\mu,\vb{q}}(\vb{k}-\vb{q}) \psi_{\phi_p',\phi_h}^{\nu,\vb{q}+\vb{\delta}_a}(\vb{k}-\vb{q}) \braket{u_{\phi_p}(\vb{k})}{u_{\phi_p'}(\vb{k}+\vb{\delta}_a)}
\end{equation}
and
\begin{equation}
    t_{h,a}^{\mu\nu}(\vb{q})
    =
    \sum_{\vb{k}} \sum_{\phi_p} \sum_{\phi_h,\phi_h'}
    \bar{\psi}^{\mu,\vb{q}}_{\phi_h,\phi_p}(\vb{k}+\vb{\delta}_a) \psi^{\nu, \vb{q}+\vb{\delta}_a}_{\phi_h',\phi_p}(\vb{k})
    \braket{u_{\phi_h'}(\vb{k})}{u_{\phi_h}(\vb{k}+\vb{\delta}_a)},
\end{equation}
where $\vb{\delta}_a$ points along primitive reciprocal lattice vector $\vb{g}_a$ and has magnitude $|\vb{\delta}_a| = 2\pi/L.$ From these we construct the link variables
\begin{equation}
    U_a^{p/h}(\vb{q}) = \text{det}\left( t_{p/h,a}(\vb{q}) \right) / \left|\text{det}\left( t_{p/h,a}(\vb{q} \right)\right|,
\end{equation}
and the Berry curvature is given by
\begin{equation}
    \Omega^{p/h}(\vb{q}) = \frac{i}{|\vb{\delta}_1 \times \vb{\delta}_2|}\ln \left[\frac{U_1^{p/h}(\vb{q}) U_2^{p/h}(\vb{q}+\vb{\delta}_1)}{U_1^{p/h}(\vb{q}+\vb{\delta}_2) U_2^{p/h}(\vb{q})}\right].
\end{equation}


%